\documentclass[12pt]{article}
\usepackage{graphicx}
\usepackage{amsmath}
\begin{document}

\title{Q-stars and charged q-stars}

\author{Athanasios Prikas}

\date{}

\maketitle

Physics Department, National Technical University, Zografou
Campus, 157 80 Athens, Greece.\footnote{e-mail:
aprikas@central.ntua.gr}

\begin{abstract} We present the formalism of q-stars with local or
global $U(1)$ symmetry. The equations we formulate are solved
numerically and provide the main features of the soliton star. We
study its behavior when the symmetry is local in contrast to the
global case. A general result is that the soliton remains stable
and does not decay into free particles and the electrostatic
repulsion preserves it from gravitational collapse. We also
investigate the case of a q-star with non-minimal energy-momentum
tensor and find that the soliton is stable even in some cases of
collapse when the coupling to gravity is absent.
\end{abstract}

PACS number(s): 11.27.+d, 04.40.-b

\newpage

\section{Introduction}

Boson stars were first discussed in the work of Kaup and Ruffini
and Bonazzola \cite{1,2}. They investigated the case of a
massive, free, complex scalar field. Complex scalar fields with a
quartic self-interaction have been studied in \cite{3,4}. The
case of charged boson stars was thoroughly investigated in
\cite{5}. The general result is that these boson stars are rather
small, with radius of order of the inverse Compton wavelength.
Gravity plays a stabilizing role. Another sort of boson stars are
the soliton stars. Their feature is that in the absence of
gravity they remain stable, as simple solitonic solutions to
highly non-linear Euler-Lagrange equations of scalar fields
\cite{6,7}. These are very large solitons with typical radius
$\sim M_{Pl}^2/{\phi}_0^3$ and mass $\sim M^4_{Pl}/{\phi}_0^3$,
where ${\phi}_0$ is a typical value of the scalar field. These
soliton stars resulted from non-topological solitons, earlier
discussed in \cite{8}. Another kind of non-topological solitons
have been studied in the literature called q-balls \cite{9}. They
appear at the minimum of the ${(U/{|\phi|}^2)}^{1/2}$ quantity in
Lagrangians with a global $U(1)$ symmetry. In non-topological
solitons discussed in \cite{6,7}, the fields rotate in their
internal $U(1)$ symmetry space with an angular velocity $\omega$
which is very small compared to the mass of the free particles.
So, the "kinetic" energy resulting from the temporal variation of
the field can be neglected. This is not the case for q-balls or
q-stars where this frequency is of the same order as the mass of
the free particles, because it is related to the parameters of
the potential (mass, trilinear parameters etc.) The surface of
the non-topological soliton stars studied in \cite{8,9} contains
an amount of energy which is of the same order as the energy
stored in the interior. In contrast, the q-star surface contains
a negligible amount of energy, because its thickness is of order
of the inverse Compton wavelength. This condition gives a
different relation for the total soliton mass and its radius, as
we shall see. Q-stars with one and two bosons in the global case
of the symmetry have been investigated in \cite{15}. Also,
q-stars including a boson and one or more fermions have been
studied in \cite{16,17,18,19}, mainly as a possible descriptive
model for neutron stars.

We expect to observe q-balls in supersymmetric extensions of
Standard Model \cite{10,11,12}, provided that the scalar
potential decreases firstly slower than ${|\phi|}^2$ and then
faster and there is a conserved quantum number concerning scalar
supersymmetric partners of fermions. If this supersymmetric
theory is an effective, low energy one, approximating a more
general, probably superstring theory in more than four
dimensions, we generally expect the scalar field to couple to
gravity in a non-trivial way, usually with a ${|\phi|}^2R$ term,
\cite{13,14}, where $R$ is the curvature scalar. Also, in a more
realistic theoretical framework, scalar fields may not be
governed by a global symmetry but by a local one. So, it would be
interesting to investigate the possibility of having charged
q-stars and to compare their properties when varying the field
strength or the frequency, which will be proved very important
for the formation of the field configuration.

\section{The general framework}

We consider a static, spherically symmetric metric:
\begin{equation}\label{1}
ds^2=e^{\nu}dt^2-e^{\lambda}dr^2-r^2d{\alpha}^2-r^2{\sin}^2\alpha
d{\beta}^2
\end{equation}
where $\nu\equiv\nu(r)$ and the convention $g_{00}=e^{\nu}$
holds. The action in natural units for a general scalar field
with a local $U(1)$ symmetry coupled to gravity is
\begin{equation}\label{2}
S=\int d^4x\sqrt{-g}\left[ \frac{R}{16\pi G}
+g^{\mu\nu}{(D_{\mu}\varphi)}^{\ast}(D_{\nu}\varphi)-U-\frac{1}{4}F_{\mu\nu}F^{\mu\nu}
\right]
\end{equation}
where
$$F_{\mu\nu} \equiv
{\partial}_{\mu}A_{\nu}-{\partial}_{\nu}A_{\mu} \hspace{3em}
D_{\mu}\varphi \equiv{\partial}_{\mu}\varphi+\imath
eA_{\mu}\varphi.$$ where $e$ is the charge (or the field
strength). Einstein equations are
\begin{equation}\label{3}
R^{\mu}_{\nu}-\frac{1}{2}{\delta}_{\nu}^{\mu}R=8\pi GT^{\mu}_{\nu}
\end{equation}
with the energy-momentum tensor:
\begin{eqnarray}\label{4}
T_{\mu\nu}={(D_{\mu}\varphi)}^{\ast}(D_{\nu}\varphi)+(D_{\mu}\varphi){(D_{\nu}\varphi)}^{\ast}-
g_{\mu\nu}[g^{\alpha\beta}{(D_{\alpha}\varphi)}^{\ast}(D_{\beta}\varphi)]
\nonumber\\ +g_{\mu\nu}U+
 \frac{1}{4}g_{\mu\nu}F^{\alpha\beta}F_{\alpha\beta}-g^{\alpha\beta}F_{\mu\alpha}F_{\nu\beta}.
\end{eqnarray}
The Euler-Lagrange equation for the matter field is
\begin{equation}\label{5}
\left(1/(\sqrt{|g|})D_{\mu}(\sqrt{|g|}g^{\mu\nu}D_{\mu})
+\frac{dU}{d{|\varphi|}^2}\right)\varphi=0.
\end{equation}
There is a Noether charge:
\begin{eqnarray}\label{6}
j^{\mu}=\sqrt{-g}g^{\mu\nu}\imath({\varphi}^{\ast}D_{\mu}\varphi-\varphi
D_{\mu}{\varphi}^{\ast})= \nonumber\\
\sqrt{-g}g^{\mu\nu}\imath(
{\varphi}^{\ast}{\partial}_{\nu}\varphi-\varphi{\partial}_{\nu}{\varphi}^{\ast}+
2eA_{\nu}{|\varphi|}^2).
\end{eqnarray}
This charge is conserved according to the equation:
\begin{equation}\label{7}
j^{\mu}_{\ ;\mu}=0.
\end{equation}
The total charge is defined:
\begin{equation}\label{8}
Q=\int d^3xj^0.
\end{equation}

We want to study static solutions, which means that the metric
and the energy-momentum tensor must be time-independent, though
the matter field, $\varphi$, may depend on time. A plausible
ansatz is to choose
\begin{equation}\label{9}
\varphi=\sigma e^{\imath \omega t}.
\end{equation}
This is the ansatz used in order to take q-ball type solutions.
Specifically, this type of non-topological solitons is stable
when $\sigma$ is a step-function and
$\omega={\omega}_{crit}\equiv {(U/{\varphi}^2)}^{1/2}_{min}$. We
also choose $A_{\mu}=(A_0,0,0,0)$ in order not to have magnetic
fields and in accordance with the static character of the
configuration. As a simple self-consistency check, we can easily
find that the energy-momentum tensor, and the metrics
consequently, is time-independent. With these substitutions the
Einstein equations are
\begin{eqnarray}\label{10}
{\lambda}'=(1-e^{\lambda})/r+8\pi Gr^2e^{\lambda} \nonumber\\
\left[{(\omega+eA_0)}^2e^{-\nu}{\sigma}^2+U+{\sigma
'}^2e^{-\lambda}+(1/2){A'}_0^2 e^{-\nu -\lambda}\right],
\end{eqnarray}
\begin{eqnarray}\label{11}
{\nu}'=(e^{\lambda}-1)/r+8\pi Gr^2e^{\lambda} \nonumber\\
\left[{(\omega+eA_0)}^2e^{-\nu}{\sigma}^2-U+{\sigma
'}^2e^{-\lambda}-(1/2){A'}_0^2 e^{-\nu -\lambda}\right].
\end{eqnarray}
and the Euler-Lagrange equations for the scalar and the gauge
field are
\begin{equation}\label{12}
{\sigma}''+[2/r+(1/2)({\nu}'-{\lambda}')]{\sigma}'+e^{\lambda}
{(\omega+eA_0)}^2e^{-\nu} \sigma
-e^{\lambda}\frac{dU}{d{\sigma}^2}\sigma=0,
\end{equation}
\begin{equation}\label{13}
A_0''+[2/r-(1/2)({\nu}'+{\lambda}')]A_0'-2e{\sigma}^2e^{\lambda}(\omega+eA_0).
\end{equation}

We will now give the formulas for the energy and charge of the
configuration. There are two alternative formulas for the charge:
\begin{equation}\label{14}
Q=8\pi\int_0^{\infty}drr^2(\omega+eA_0){\sigma}^2e^{(\lambda-\nu)/2},
\end{equation}
\begin{equation}\label{15}
Q=4\pi r^2A_0', \hspace{2em} r\rightarrow \infty.
\end{equation}
Also,there are two alternative formulas for the total energy (the
mass) of the soliton
\begin{equation}\label{16}
\textrm{E}=4\pi\int_0^{\infty}drr^2[{(\omega+eA_0)}^2{\sigma}^2e^{-\nu}+U+e^{-\lambda}{\sigma'}^2
+(1/2){A_0'}^2e^{-\nu-\lambda}].
\end{equation}
If we know the total charge of the soliton we can find the total
mass from the asymptotic relation:
\begin{equation}\label{17}
e^{\lambda}={(1-2GE/r+GQ^2/4\pi r^2)}^{-1}, \hspace{2em}
r\rightarrow \infty.
\end{equation}
We will mainly use eq. \ref{14}, \ref{16}.

We have formulated the general framework concerning the case of
local $U(1)$ symmetry. In order to examine the global case we put
$e=0$ and $A=0$ and the results of \cite{15} can be obtained. It
is also very convenient to define
\begin{equation}\label{18}
\theta\equiv\omega+eA_0.
\end{equation}
In the case of the global symmetry $\theta\rightarrow\omega$.

\section{The q-star solution}

For the sake of convenience we will use a simple potential of the
form
\begin{equation}\label{19}
U={\sigma}^2\left(1-{\sigma}^2+\frac{1}{3}{\sigma}^4\right).
\end{equation}
We will examine separately three different regions: The interior,
the surface and the exterior of the soliton. At the interior the
metric, gauge and matter fields vary very smoothly with respect
to the radius, because, as we can see from Einstein equations,
the metric derivatives are proportional to $8\pi G$ times
${\varphi}^4$ in rough estimate, which is a very small quantity.
Let ${\sigma}_0$ be a value of the order of magnitude of the
matter field values inside the soliton. We will make the
rescallings:
\begin{equation}\label{20}
\tilde{r}=r{\sigma}_0, \hspace{1em}
\tilde{\sigma}=\sigma/{\sigma}_0, \hspace{1em}
\tilde{\theta}=\theta/{\sigma}_0, \hspace{1em}
\tilde{U}=U/{\sigma}_0^4.
\end{equation}
We also define
\begin{equation}\label{21}
B\equiv e^{-\nu}, \hspace{1.4em} A\equiv e^{-\lambda}.
\end{equation}

The idea which the q-ball type soliton rely on, in contrast to
other sorts of non-topological solitons, is the relation:
$$\omega \sim m\sim\varphi$$
The first part of the condition comes from the relation combining
${\omega}_{crit}$ to the potential parameters (i.e.
${\omega}_{crit}={(U/{|\varphi|}^2)}_{min}^{1/2}$). The second is
derived from solving the equation of motion of the matter field,
so as to find its value. Remembering that the gravity becomes
important when $R\sim GM$, where $M$ the soliton mass, we can
easily find that the star radius is $R \sim M_{Pl}/{\sigma}_0^2$.
We now define the quantity:
\begin{equation}\label{22}
\epsilon=\sqrt{8\pi G}{\sigma}_0\sim m/M_{Pl}.
\end{equation}
This means that $\epsilon$ is very small. It is easy to find that
the radius of the soliton is the same order of magnitude as
${\epsilon}^{-1}$, or, to be more definite, if ${\tilde{r}}_{-}$
is the radius of the interior of the soliton, then for the
interior holds
\begin{equation}\label{23}
\tilde{r}={\tilde{r}}_{-}x={\epsilon}^{-1}\kappa x, \hspace{1.2em}
0\leq x\leq 1
\end{equation}
where $\kappa$ is of order unity. We also define
\begin{equation}\label{24}
{\tilde{e}}^2=e^2{({\epsilon}^{-1}\kappa)}^2.
\end{equation}
So, for the interior, dropping the tildes and the $0(\epsilon)$
terms the Einstein equations are
\begin{equation}\label{25}
1-A-x\frac{dA}{dx}=x^2{\kappa}^2\left[{\theta}^2{\sigma}^2B+U+
\frac{1}{2e^2}{\left(\frac{d\theta}{dx}\right)}^2AB\right],
\end{equation}
\begin{equation}\label{26}
A-1-x\frac{A}{B}\frac{dB}{dx}=x^2{\kappa}^2\left[{\theta}^2{\sigma}^2B-U-\frac{1}{2e^2}
{\left(\frac{d\theta}{dx}\right)}^2AB\right].
\end{equation}
which result from the $G_0^0=8\pi GT_0^0$ and $G_1^1=8\pi GT_1^1$
respectively, and the Euler-Lagrange equations:
\begin{equation}\label{27}
{\theta}^2B-\frac{dU}{d{\sigma}^2}=0,
\end{equation}
\begin{equation}\label{28}
\frac{d^2\theta}{dx^2}+\left[\frac{2}{x}+\frac{1}{2}\left(\frac{1}{A}
\frac{dA}{dx}+\frac{1}{B}\frac{dB}{dx}\right)\right]\frac{d\theta}{dx}
-\frac{2e^2{\sigma}^2{\theta}}{A}=0.
\end{equation}
These equations are in a manageable form because the derivatives
of the metrics with respect to $x$ are not too small as those
with respect to $r$ and the soliton radius in units of
${{\sigma}_0}^{-1}$ is not "huge" any more. It is a matter of
simple algebra to verify that eq. \ref{27} has the correct limit
when we study the global case (i.e.: $\theta \rightarrow \omega$)
and the gravity is absent (so $B\rightarrow1$).

The boundary conditions concerning the soliton interior are
$A(0)=1$, $A(1)=1/B(1)$ and $\theta'(0)=0$. The first condition
results from the freedom to rescale the metrics, the second from
the fact that outside the soliton (we will see that the surface
region is extremely thin, so one can define as "matter region"
the soliton interior) the solution to the Einstein equations
should be Schwarzschild and the third condition reflects the
absence of electric field at the center of the soliton.

Eq. \ref{27} can provide an exact solution for the matter field,
i.e.:
\begin{equation}\label{29}
\sigma=\sqrt{1+\theta\sqrt{B}}.
\end{equation}
Then:
\begin{equation}\label{30}
U=\frac{1}{3}(1+{\theta}^3B^{3/2}).
\end{equation}

The surface is very thin in the case of q-stars and this is a
crucial difference between them and other non-topological-soliton
stars. The surface thickness is of order ${\sigma}^{-1}$. In this
type of non-topological solitons $\sigma\sim\omega$, in contrast
to other non-topological soliton. The surface energy is of order
${\sigma}^4R^2{\sigma}^{-1}$. We saw that $R\sim
M_{Pl}/{\sigma}^2$. The energy stored in the soliton interior is
$E_{int}\sim R^3{\sigma}^4$. So the relation: $E_{sur}\sim
\epsilon E_{int}$ hold. These results are numerically verified.
So the surface energy is negligible, a result that not hold for
the other kind of soliton stars. Within the surface region, the
metric and the gauge field varies very slowly (their derivatives
are of $0({\epsilon}^{-1})$) and can be regarded as constant. The
matter field varies rapidly (this is the meaning of a surface).
Now we will find the "eigenvalue" equation for the gauge field,
or the frequency in the global case. In eq. \ref{12}
$2/r+(1/2)({\nu}'-{\lambda}')$ are $0(\epsilon)$ terms and can be
neglected. Eq. \ref{12} can be straightforward integrated giving
as result:
\begin{equation}\label{31}
e^{-\lambda}{\left(\frac{d\sigma}{dr}\right)}^2+{\theta}^2{\sigma}^2e^{-\nu}-U=0.
\end{equation}
In the interior the above equation, neglecting the $0(\epsilon)$
terms, will contain the second and third term. In order to confirm
the continuity of the above expression (and should be continuous
as a first integral) we should set
\begin{equation}\label{32}
{\left({\theta}^2{\sigma}^2e^{-\nu}-U\right)}_{r_{-}}=0.
\end{equation}
Switching off both gravity and the gauge field (i.e. making the
symmetry global) the above expression arises from the
Euler-Lagrange equation when the spatial derivative terms are
neglected (i.e. the soliton is very large) and
$\omega={\omega}_{min}$. Now $\theta$ took the place of $\omega$.
Eq. \ref{32} gives the eigenvalue condition for the frequency.
Substituting equations \ref{29}, \ref{30} into eq. \ref{32} we
find that for the surface
\begin{equation}\label{33}
{\theta}^2_{sur}B_{sur}=1/4.
\end{equation}
We can easily understand that different values of the potential
parameters gives a different relation between $\theta$ and the
metrics. When both gravity and electric charge are absent the
above relation takes the form $\omega=1/2$, which is really the
value ${(U/{|\varphi|}^2)}^{1/2}_{min}$ for the potential of eq.
\ref{19}. Eq. \ref{33} gives a measure of the gravity strength
when we know the frequency. When $A(sur)<1$, i.e. when we switch
on gravity, the frequency of the large soliton (thin-wall
approximation) is less than that corresponding to the absence of
gravity. This fact summarizes the main effect of the gravity upon
the soliton properties. The "potential" that enters the
${\omega}_{min}$ relation now is not $U$ but $A(sur)U$ and is
smaller than in the case that the gravity is absent.

In the exterior region the matter field $\sigma$ is zero. The
metric fields should obey the asymptotic relation $$A(r)=1/B(r),
\hspace{2em} r\rightarrow\infty$$ One can prove that the above
condition holds for the entire exterior region, not only at
infinity, using equations
$$1-A-x\frac{dA}{dx}=x^2{\kappa}^2
\frac{1}{2e^2}{\left(\frac{d\theta}{dx}\right)}^2AB,$$
$$A-1-x\frac{A}{B}\frac{dB}{dx}=-x^2{\kappa}^2\frac{1}{2e^2}
{\left(\frac{d\theta}{dx}\right)}^2AB, \hspace{1em} x>1$$ which
concern the metrics outside the soliton, and the above boundary
condition. This result is in agreement with the Schwarzschild
type of the exterior solution. The boundary condition for the
gauge field $\theta$ results from the matching of the exterior
end the interior solutions. The other condition for the gauge
field is $\theta\rightarrow\omega$ when $r\rightarrow\infty$
which means that for large distances the "true" gauge field,
$(\theta-\omega)/e$, tends to zero. Our results also hold in the
global case of the $U(1)$ symmetry studied in \cite{15}.

\section{Numerical results}

The computation of energy (total mass) and particle number of the
field configuration are based on eq. \ref{14}, \ref{16}. $\kappa$
gives a measure of the soliton radius. The main features of the
soliton are plotted as functions of the frequency, which is the
most crucial parameter. Alternatively, we could use a parameter of
the surface gravity ($A$ or $1/B$) in order to test the soliton
behavior. But ${\theta}_{sur}$, or $\omega$, are not independent
parameters, as we can see from equation \ref{33}. So, the value
of the frequency gives a measure of the surface gravity strength
as well.

\begin{figure}
\centering
\includegraphics{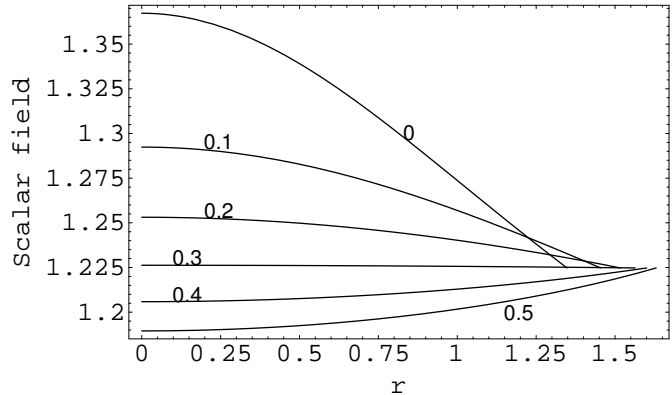}
\caption{Soliton profiles for different values of the coupling
constant, with $A(sur)=0.36$ or, equivalently,
${\theta}_{sur}=0.3$ for five different values of the field
strength $e$. We examine only the interior region because the
surface is very thin.} \label{figure1a}
\end{figure}

The total particle number can be identified with the total charge
if we take as unity the charge of each particle. Figure
\ref{figure1a} shows the behavior of the soliton profile when
decreasing slowly the field strength $e$. We sketched only the
interior region because the surface is very thin and the field
varies rapidly. When the coupling constant increases, the soliton
gets larger due to the electrostatic repulsion and, most
interesting, the value at the origin is not the maximum one, due
to the same effect.

\begin{figure}
\centering
\includegraphics{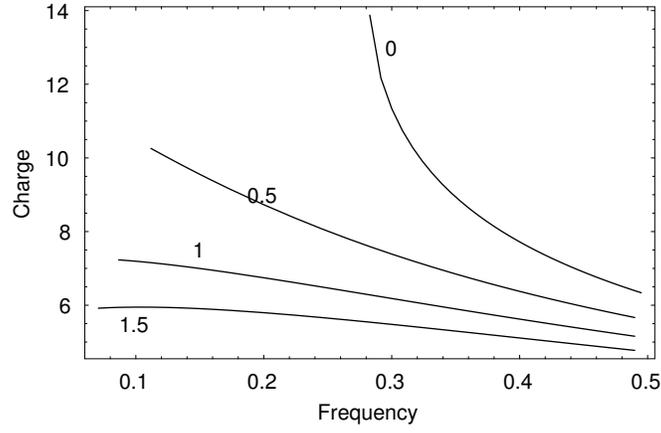}
\caption{The total charge (particle number) of the soliton as a
function of ${\theta}_{sur}$, which equals to $\omega$ in the
global case, for four different values of the field strength.}
\label{figure1}
\end{figure}

Fig.\ref{figure1} shows the behavior of the soliton charge as a
function of the frequency. These results are intuitively expected.
When ${\theta}_{sur}$ or $\omega$ decreases, the particle number
increases because the gravity gets more important (eq. \ref{33}).
The presence of the electric charge $e$ makes the decrease of the
particle number slower, due to the electrostatic repulsion between
the different parts of the soliton.

\begin{figure}
\centering
\includegraphics{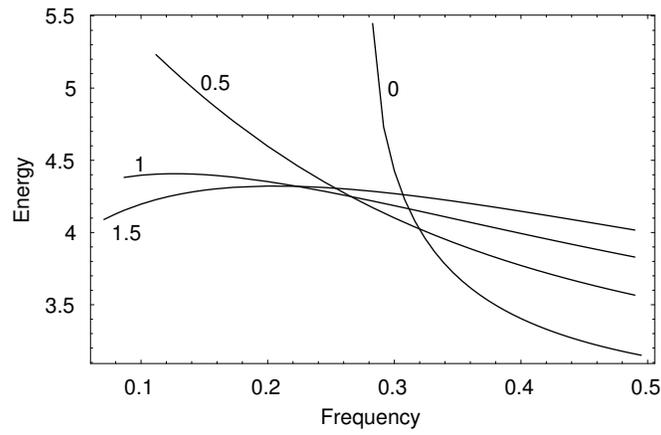}
\caption{The mass of the field configuration as a function of
${\theta}_{sur}$ for four different values of the field
strength.} \label{figure2}
\end{figure}

The results of figure \ref{figure2} are more interesting. In the
case of global symmetry, the increase of frequency decreases the
total soliton energy due to the decrease of the particle number.
When the field strength $e$ is large, the decrease of the particle
number is not enough to decrease the energy and the electrostatic
energy should be taken into account. This energy becomes more
important when gravity weakens, i.e. when frequency approaches
the value that has when gravity is absent. For the potential of
eq. \ref{19} this value is ${(U/{|\varphi|}^2)}^{1/2}_{min}=0.5$.

\begin{figure}
\centering
\includegraphics{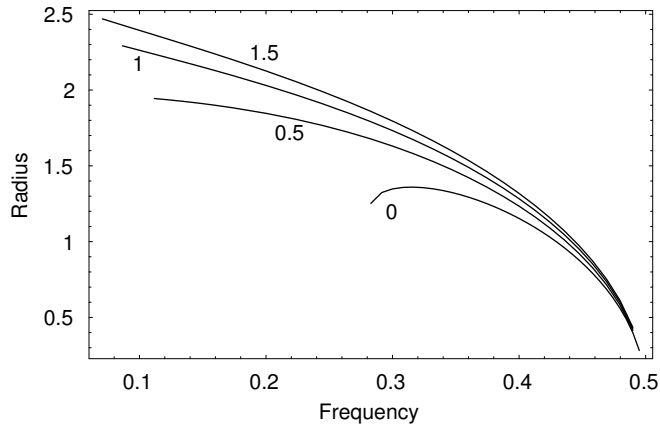}
\caption{The soliton radius as a function of ${\theta}_{sur}$ for
four different values of the field strength.} \label{figure3}
\end{figure}

\begin{figure}
\centering
\includegraphics{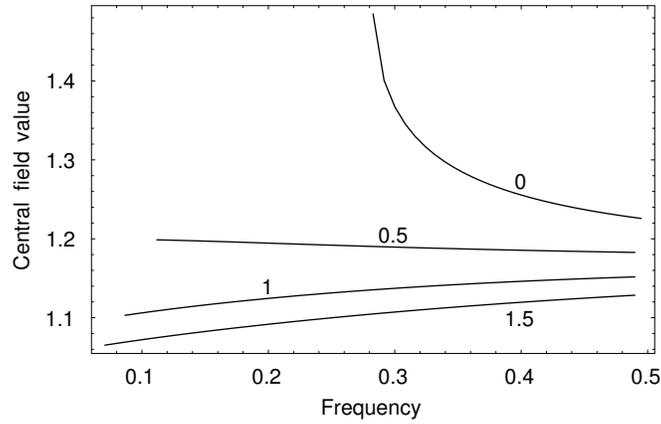}
\caption{The value of the scalar field at center of the soliton as
a function of ${\theta}_{sur}$.} \label{figure4}
\end{figure}

The soliton radius, figure \ref{figure3}, has a very interesting
behavior. In general the radius gets larger with the coupling
constant due to the electrostatic repulsion between the different
parts of the soliton. The radius also increases when the
frequency decreases, because the particle number gets larger. But
in the global case and after a certain point the soliton radius
decreases because the gravity becomes extremely strong. This is
the boundary between stars and black holes, because when
$\omega\rightarrow 0$, then $A(sur)\rightarrow 0$ (eq. \ref{33}
and the relation $A(sur)=1/B(sur)$ between the metrics) and the
metric $B$ at the center of the soliton approaches infinity.

\begin{figure}
\centering
\includegraphics{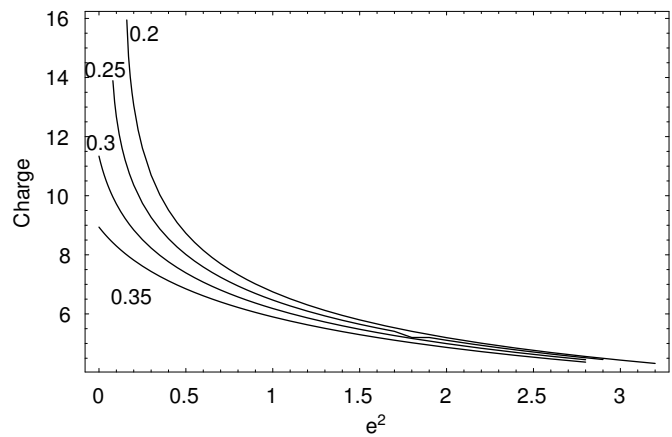}
\caption{The soliton charge as a function of the coupling
constant for four different frequencies.} \label{figure5}
\end{figure}

\begin{figure}
\centering
\includegraphics{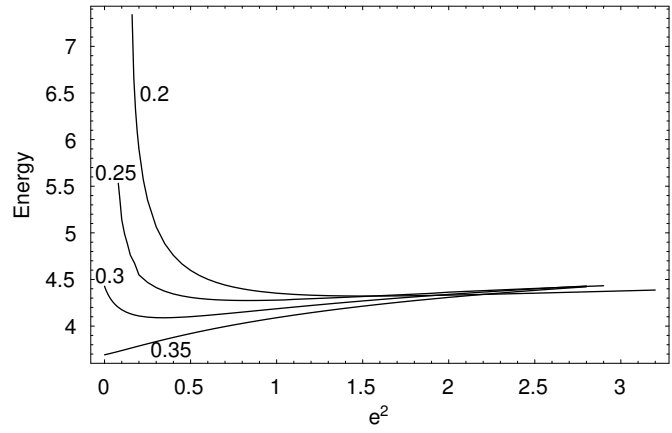}
\caption{The total soliton mass as a function of the coupling
constant $e$.} \label{figure6}
\end{figure}

\begin{figure}
\centering
\includegraphics{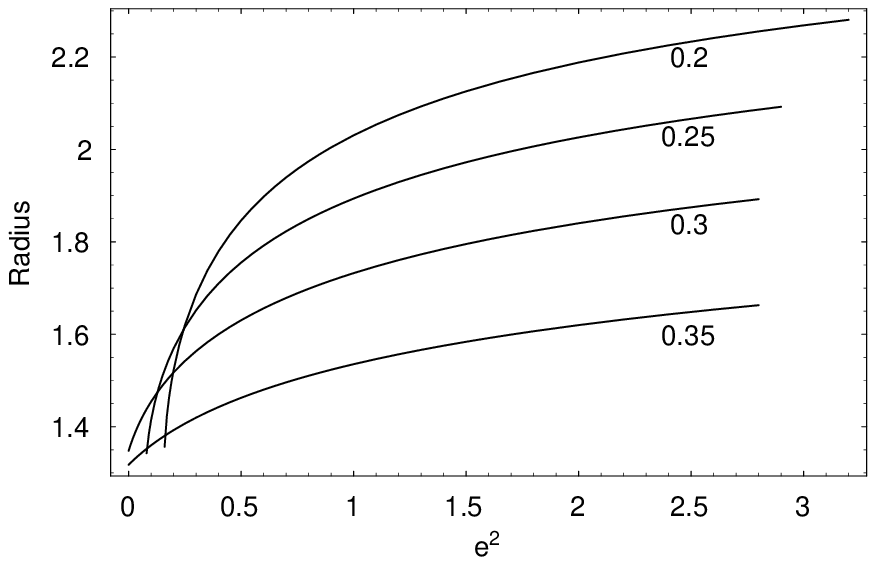}
\caption{The soliton radius as a function of the coupling
constant.} \label{figure7}
\end{figure}

We will now study the soliton behavior when varying the coupling
constant $e$. A general feature is the decrease of the soliton
mass, radius and particle number when the coupling constant gets
larger. We also observe the increase in the ratio soliton energy
per particle number (which expresses the ratio soliton energy per
free particles energy) which leads to an "electrostatic"
instability of the soliton for very large values of the field
strength $e$. The field configurations described in the figures
satisfy $E_{sol}/E_{free}<1$, i.e. they are stable
configurations. In order to simplify calculations we chose the
mass of the free particles to be 1 (eq. \ref{19}), so the energy
of the free particles with the same charge as the soliton equals
the particle number. When the above ratio exceeds unity, the
soliton decays into free particles. On the other hand the
presence of non-zero electric charge prevents soliton from
gravitational collapse. In the global case and when $\omega$ is
very small, the metric $B$ at the center of the soliton tends to
infinity, revealing a singularity at the center of the field
configuration. Electrostatic repulsion removes this singularity
and makes the soliton stable, even for lower frequencies.

The plot of charge as a function of the field strength, fig.
\ref{figure5}, has a well-understood behavior. The increase in the
electrostatic repulsion decreases the number of particles within
the soliton. The plot of energy, fig. \ref{figure6}, is more
interesting. When the frequency is small and, consequently the
gravity is more important (eq. \ref{33}) the increase of the
coupling constant leads to a decrease of the energy due to the
decrease in the total charge. When gravity is weaker (for
instance $\theta=0.35$ or, equivalently $A(sur)=0.49$), the
increase of the coupling constant increases the electrostatic
energy. The behavior of the soliton radius is interesting as
well. Large coupling constant leads to bigger solitons when the
strengthening of gravity has the opposite effect.

\section{Q-stars with non-minimal energy-\\ momentum tensor}

We will now examine the case where the action has a new term
concerning the coupling of the scalar field to gravity. Such
terms can arise in effective Lagrangians when considering
theories in more than four dimensions. We will use a simple, toy
model for the action, namely:
\begin{equation}\label{34}
S=\int d^4x \sqrt{-g}\left[\left(\frac{1}{16\pi G} +\xi
\varphi{\varphi}^{\ast}\right) R+g^{\mu\nu}{\partial}_{\mu}\varphi
{\partial}_{\nu}{\varphi}^{\ast}-U\right].
\end{equation}
The energy-momentum tensor is
\begin{eqnarray}\label{35}
T_{\mu\nu}={\partial}_{\mu}{\varphi}^{\ast}{\partial}_{\nu}\varphi+
{\partial}_{\nu}{\varphi}^{\ast}{\partial}_{\mu}\varphi-
g_{\mu\nu}(g^{\alpha\beta}{\partial}_{\alpha}{\varphi}^{\ast}
{\partial}_{\beta}\varphi)+g_{\mu\nu}U+ \nonumber\\
2\xi{\varphi}^{\ast}\varphi(R_{\mu\nu}-\frac{1}{2}g_{\mu\nu}R)
-2\xi
g_{\mu\nu}g^{\alpha\beta}{\partial}_{\alpha}{\partial}_{\beta}
{|\varphi|}^2+2\xi{\partial}_{\mu}{\partial}_{\nu}{|\varphi|}^2.
\end{eqnarray}
We have regarded only the case of global $U(1)$ symmetry for
simplicity.

Using the ansatz of eq. \ref{9} referring to the static
configurations we take the Euler-Lagrange equation for the matter
field:
\begin{equation}\label{36}
{\sigma}''+\left[\frac{2}{r}+\frac{1}{2}({\nu}'-{\lambda}')\right]
+e^{\lambda-\nu}{\omega}^2{\sigma}+e^{\lambda}\xi R\sigma-
e^{\lambda}\frac{dU}{d{\sigma}^2}\sigma=0,
\end{equation}
and the Einstein equations:
\begin{eqnarray}\label{37}
(1+16\xi\pi
G)\left(\frac{\lambda'}{r}+\frac{e^{\lambda}}{r^2}-\frac{1}{r^2}\right)
=8\pi G[e^{\lambda-\nu}{\omega}^2{\sigma}^2+U+ \nonumber\\
{\sigma'}^2(1+4\xi) -2\xi\nu'
\sigma'\sigma+4e^{\lambda}\xi(U-{\omega}^2{\sigma}^2e^{-\nu} -\xi
R{\sigma}^2)],
\end{eqnarray}
\begin{eqnarray}\label{38}
(1+16\xi\pi
G)\left(\frac{\nu'}{r}-\frac{e^{\lambda}}{r^2}+\frac{1}{r^2}\right)
=8\pi G[e^{\lambda-\nu}{\omega}^2{\sigma}^2- \nonumber\\
U+{\sigma'}^2-2\xi\sigma\sigma'(\nu'+4/r)],
\end{eqnarray}
\begin{eqnarray}\label{39}
[1+16\xi\pi G{\sigma}^2(1+6\xi)]R= \nonumber\\ 8\pi G
[(2+12\xi)(e^{-\lambda}{\sigma'}^2-e^{-\nu}{\omega}^2{\sigma}^2)+4U(1+3\xi)],
\end{eqnarray}
where the last equation has been obtained by taking the trace of
the Einstein equations and will be used for the calculation of the
curvature scalar.

The total mass of the configuration is
\begin{eqnarray}\label{40}
\textrm{E}=4\pi\int_0^{\infty}drr^2\frac{1}{1+16\xi\pi
G{\sigma}^2} [e^{-\nu}{\omega}^2{\sigma}^2(1-4\xi)+ \nonumber\\
U(1+4\xi)+e^{-\lambda}{\sigma'}^2(1+4\xi)
-2e^{-\lambda}\xi\nu'\sigma'\sigma-4{\xi}^2\sigma R].
\end{eqnarray}
The total particle number is given by eq. \ref{5}, or,
equivalently, \ref{14}. The potential is the same as in eq.
\ref{19}. We also make the rescallings and the definitions
\ref{20} and \ref{21}. We also define
\begin{equation}\label{41}
\tilde{R}=R/{\sigma}_0^2.
\end{equation}
We discriminate between three regions: The interior region, the
very thin surface region and the exterior, where the metric is
Schwarzschild. Following the steps of the previous case, we write
the equations for the interior, dropping the tildes and the
$0(\epsilon)$ quantities:
\begin{equation}\label{42}
1-A-x\frac{dA}{dx}=x^2{\kappa}^2\left[{\theta}^2{\sigma}^2B(1-4\xi)+U(1+4\xi)\right],
\end{equation}
\begin{equation}\label{43}
A-1-\frac{A}{B}\frac{dB}{dx}=x^2{\kappa}^2\left[{\theta}^2{\sigma}^2B-U\right],
\end{equation}
resulting from the $G_0^0$ and $G_1^1$ Einstein tensors
respectively. The rescalled curvature scalar, $\tilde{R}$ is of
$0(\epsilon)$ order. So the total mass of the configuration is
\begin{equation}\label{44}
\textrm{E}=4\pi\int_0^{\infty}drr^2({\theta}^2{\sigma}^2B(1-4\xi)+U(1+4\xi)).
\end{equation}
We also find that eq. \ref{29} for the field value in the soliton
interior remains valid.

We solve numerically eq. \ref{42}, \ref{43} and we find the mass
and particle number of the soliton using eq. \ref{44} and
\ref{14} respectively. Some general features of the solutions for
the $\xi\neq 0$ cases are: a)For small values of the parameter
that describes the strength of the coupling of the scalar field to
gravity the soliton becomes more "stable" against gravitational
collapse. We can find solutions for some values of the frequency
that have $B(0)\rightarrow\infty$ when $\xi$ is zero. These
solutions describe a real singularity at the soliton origin,
which is removed when we "switch on" the coupling to gravity. But
this coupling, however strong, can not prevent soliton from
gravitational collapse when $A(sur)\rightarrow 0$. b)Larger values
of the coupling parameter correspond to solutions with lower
energy and charge but larger radius, which means that the
coupling of the matter field to gravity has an effect similar to
the electrostatic repulsion. This effect has to do with the
repulsive character of the $\xi{\varphi}^2R$ term.

\begin{figure}
\centering
\includegraphics{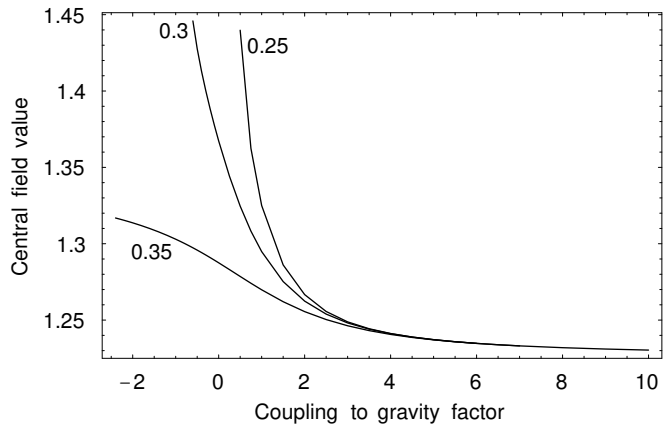}
\caption{The central field value as a function of the coupling
$\xi$ of the scalar field to gravity for three different
frequencies.} \label{figure8}
\end{figure}

\begin{figure}
\centering
\includegraphics{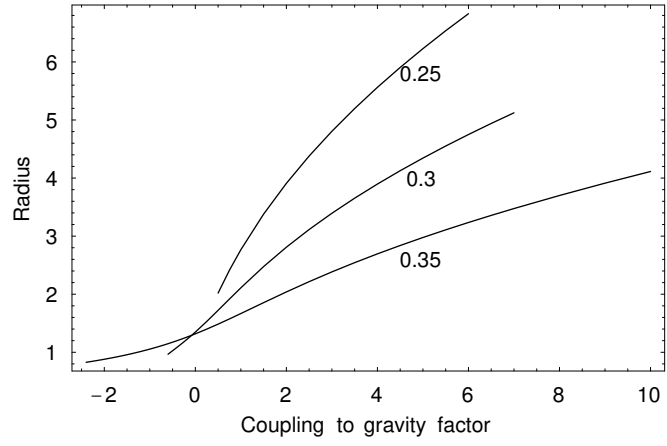}
\caption{The radius of the soliton as a function of the coupling
$\xi$ for three different frequencies.} \label{figure9}
\end{figure}

\begin{figure}
\centering
\includegraphics{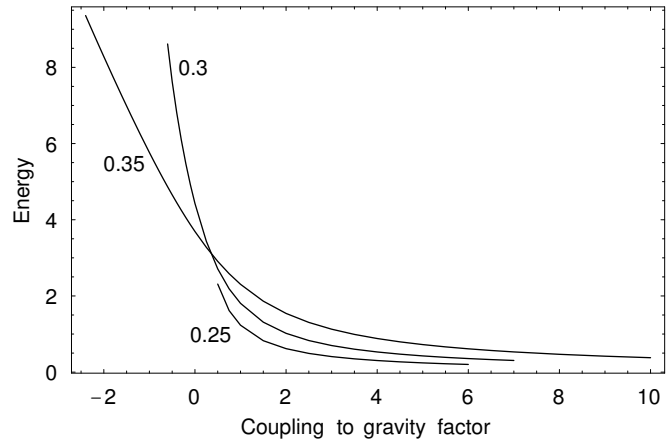}
\caption{The soliton mass as a function of $\xi$.}
\label{figure10}
\end{figure}

\begin{figure}
\centering
\includegraphics{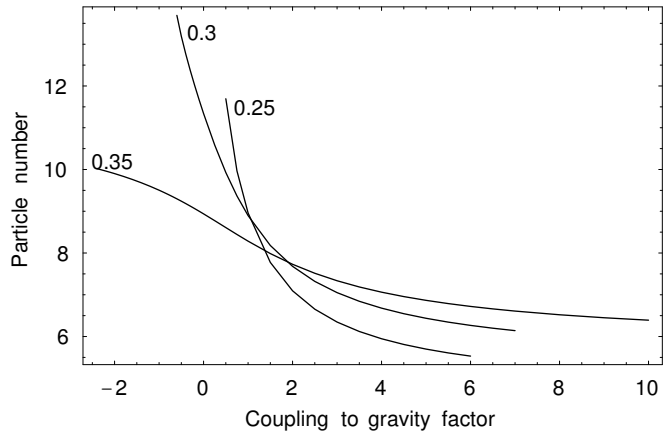}
\caption{The soliton particle number as a function of $\xi$.}
\label{figure11}
\end{figure}

\newpage

\textbf{Acknowledgments}

\vspace{1em}

I wish to thank N.D. Tracas and E. Papantonopoulos for helpful
discussions.

\end{document}